\documentclass[12pt]{article}
\usepackage{a4wide}
\usepackage{amssymb}
\usepackage{amsmath}
\usepackage{graphicx}
\usepackage{mathdots}
\usepackage{slashed}
\usepackage{verbatim}
\usepackage{xcolor}
\usepackage{cancel}

\usepackage{cite}

\usepackage{comment}

\usepackage{hyperref}
\usepackage{cleveref} 

\usepackage{IEEEtrantools}

\def\makeatletter{\catcode`\@=11}
\makeatletter
\def\mathbox#1{\hbox{$\m@th#1$}}%
\def\math@ccstyles#1#2#3#4#5#6#7{{\leavevmode
      \setbox0\mathbox{#6#7}%
      \setbox2\mathbox{#4#5}%
      \dimen@ #3%
      \baselineskip\z@\lineskiplimit#1\lineskip\z@
      \vbox{\ialign{##\crcr
             \hfil \kern #2\box2 \hfil\crcr
             \noalign{\kern\dimen@}%
             \hfil\box0\hfil\crcr}}}}
\def\mathaccstyles{\math@ccstyles\maxdimen}
\def\maththroughstyles{\math@ccstyles{-\maxdimen}}
\def\unity%
 {\maththroughstyles{.45\ht0}\z@\displaystyle {\mathchar"006C}\displaystyle 1}

\numberwithin{equation}{section}
 
\begin{document}

\mbox{}
\vspace{0truecm}
\linespread{1.1}

\def\LL{\mathcal{L}}
\def\EE{\mathcal{E}}


\centerline{\Large \bf Higher-Order Fermion Interactions in Effective}
\bigskip

\centerline{\Large \bf  Field Theories for Phase Transitions}

\vspace{.4cm}

 \centerline{\LARGE \bf }

\vspace{1.5truecm}

\centerline{
    { \bf D. Rodriguez-Gomez${}^{a,b}$} \footnote{d.rodriguez.gomez@uniovi.es}
   {\bf and}
    { \bf J. G. Russo ${}^{c,d}$} \footnote{jorge.russo@icrea.cat}}

\vspace{1cm}
\centerline{{\it ${}^a$ Department of Physics, Universidad de Oviedo}} \centerline{{\it C/ Federico Garc\'ia Lorca  18, 33007  Oviedo, Spain}}
\medskip
\centerline{{\it ${}^b$  Instituto Universitario de Ciencias y Tecnolog\'ias Espaciales de Asturias (ICTEA)}}\centerline{{\it C/~de la Independencia 13, 33004 Oviedo, Spain.}}
\medskip
\centerline{{\it ${}^c$ Instituci\'o Catalana de Recerca i Estudis Avan\c{c}ats (ICREA)}} \centerline{{\it Pg.~Lluis Companys, 23, 08010 Barcelona, Spain}}
\medskip
\centerline{{\it ${}^d$ Departament de F\' \i sica Cu\' antica i Astrof\'\i sica and Institut de Ci\`encies del Cosmos}} \centerline{{\it Universitat de Barcelona, Mart\'i Franqu\`es, 1, 08028
Barcelona, Spain }}
\vspace{1cm}
\setcounter{footnote}{0}

\centerline{\bf ABSTRACT}
\medskip

We investigate the impact of higher-order fermionic deformations 
in phase transitions analogous to those described by the Bardeen–Cooper–Schrieffer (BCS) theory.
 Focusing specifically on the  8-fermion interaction, we show that this term can have significant consequences. In certain regions of parameter space, the theory continues to exhibit second-order phase transitions with mean-field critical exponents and  the same critical temperature; however, the temperature dependence of the superconducting gap can deviate markedly from conventional BCS behavior. In other regions, the theory  exhibits first-order phase transitions. We conclude by discussing potential phenomenological applications of these theories.

\noindent 

\newpage

\tableofcontents

\section{Introduction}

\begin{figure}[h!]
\centering
\includegraphics[angle=0,scale=.3]{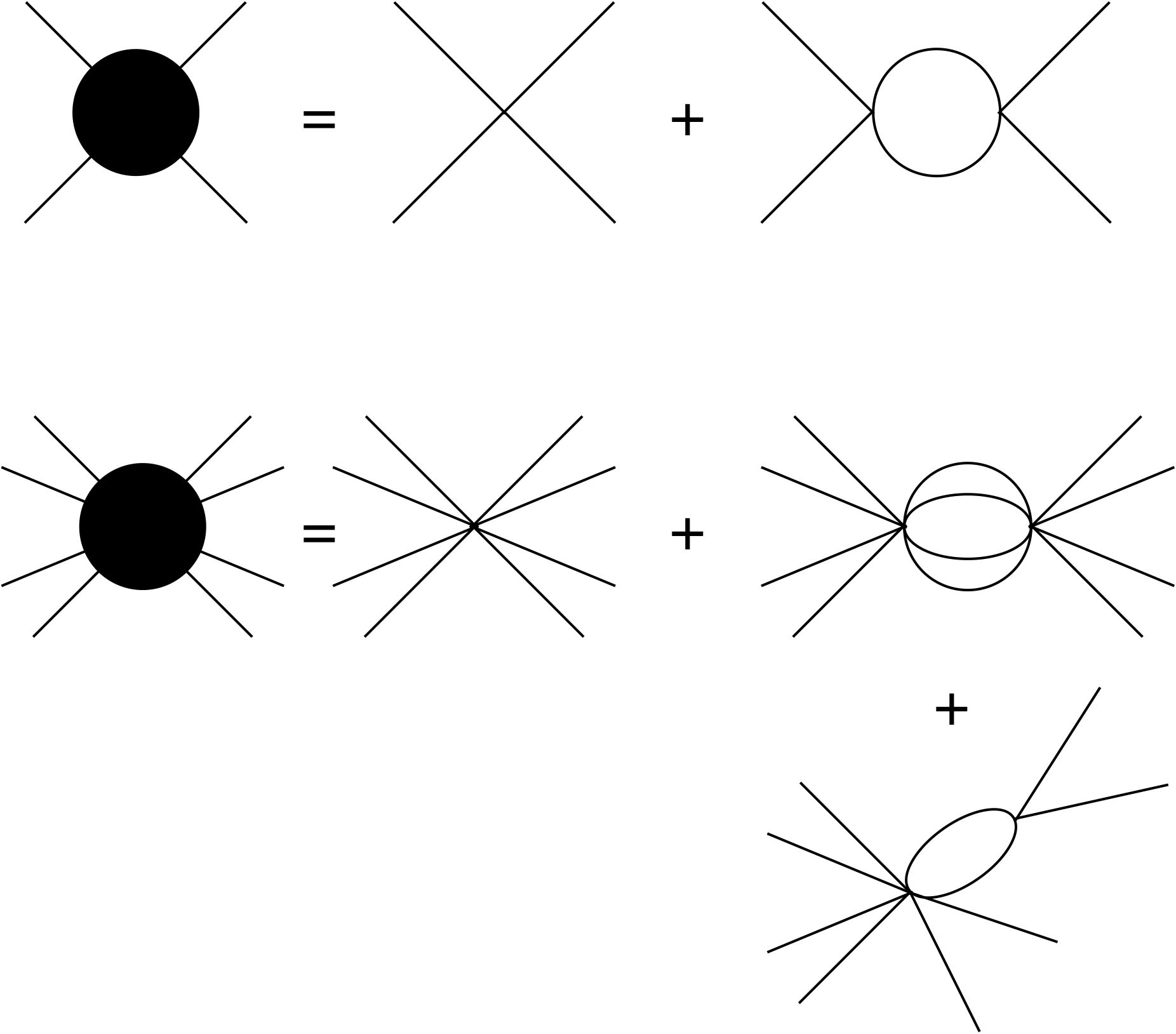}
\caption{4- and 8-point functions.}
\label{diagrams}
\end{figure}

The Bardeen-Cooper-Schrieffer (BCS) theory of superconductivity remains as a major achievement of the 20th century Physics \cite{Bardeen:1957mv}. Through phonon interactions, electrons feel an effective quartic interaction which forces them to condense into Cooper pairs driving superconductivity. 
Yet, from the perspective of effective field theory, it remains surprising that a quartic fermionic interaction can produce such a pronounced effect. Naive power-counting 
suggests that the quartic coupling is an irrelevant interaction in spacetime dimensions greater than two. However, as argued in \cite{Polchinski:1992ed}, the presence of the Fermi surface significantly modifies the naive scaling, making the quartic coupling marginally relevant in specific kinematic regimes. This observation naturally raises the question of how this effect might propagate in the presence of higher-order couplings. 
In general, the RG flow  couples the evolution of the various couplings, and it is plausible that the infrared relevance of the quartic interaction  influences the behavior of higher-order terms. To illustrate this point,  consider the following  interaction Lagrangian,
\begin{equation}
\mathcal{L}_{\rm int}=\int d^dx\, \left(g_4\,\psi^4+g_8\,\psi^8\right)\,,
\end{equation}
where $\psi$ stands for a fermionic field.\footnote{
At this stage, our discussion remains  schematic. 
The precise model will be described in  section 2.} The perturbative expansion of the 4- and 8-point function contains the Feynman diagram contributions shown in  \cref{diagrams}.
Then

\begin{equation}
G_4=g_4+g_4^2\,C\,,\qquad G_8=g_8+g_8\,g_4\,C+g_8^2\,S\,,
\end{equation}
where $C$ and $S$ represent the 1-loop and three loop diagrams. 
In a naive analysis, both $C$
 and 
$S$
 would be expected to scale with a positive power of a reference energy scale 
$E$, rendering the four-fermion and eight-fermion interactions irrelevant. In the presence of a Fermi surface, however, the scaling behavior of the diagrams is altered, and one effectively obtains \cite{Polchinski:1992ed},
\begin{equation}
C\sim C_0\,\log E\,,\qquad S\sim  S_0\,E^{a}\,,
\end{equation}
for some positive $a$, where $C_0,\,S_0$ are numerical constants. Assuming, for simplicity, a hierarchy of couplings where the $g_8^2$ term is negligible, the $\beta$-function equations for the dimensionless couplings $(\mathcal{G}_4,\,\mathcal{G}_8)=(G_4,E^aG_8)$ are of the form

\begin{equation}
E\, \partial_E\mathcal{G}_4= C_0\,\mathcal{G}_4^2\,,\qquad E\, \partial_E\mathcal{G}_8=a\,\mathcal{G}_8+C_0\,\mathcal{G}_8\,\mathcal{G}_4\, .
\end{equation}
Therefore

\begin{equation}
\mathcal{G}_4=\frac{\mathcal{G}_4^{(0)}}{1-\mathcal{G}_4^{(0)}\,C_0\,\log E}\,,\qquad \mathcal{G}_8=\frac{\mathcal{G}_8^{(0)}\,E^{a}}{1-\mathcal{G}_4^{(0)}\,C_0\,\log E}\,.
\end{equation}
Consider first the case 
$\mathcal{G}_8^{(0)}=0$, corresponding to the standard BCS scenario. It was shown in \cite{Polchinski:1992ed}
that  
$C_0<0$. As a result, the quartic coupling 
$\mathcal{G}_4$
 becomes marginally relevant and, as is well known, gives rise to the BCS theory of superconductivity. 
  When 
$\mathcal{G}_8^{(0)}\ne 0$
 is introduced, the numerator exhibits the expected power-law dependence on the reference scale characteristic of an irrelevant interaction. However, the denominator significantly modifies the RG flow: at low energies it becomes small, driving 
$\mathcal{G}_8$
 to grow toward the infrared. This schematic analysis therefore suggests that higher-order fermionic couplings may generate nontrivial infrared dynamics.


In this paper we set out to study the effect of a $\psi^8$ interaction in detail.  As we will see, depending on the coefficient of the higher-order fermionic coupling, the transition can be second order or first order.
The transition is second order below a certain critical value of the $\psi^8$ coupling. In this case, the critical exponents
still have mean-field values, but the gap curve is deformed
as compared with the BCS 
case.
In another range of parameter,  the higher-order couplings drive the system into a regime where the phase transition becomes first-order.
 We will discuss potential phenomenological applications of the various scenarios in \cref{conclusions}.

\section{The model}\label{sec:modelo}

Our starting point is  the familiar non-relativistic Lagrangian describing the BCS model, extended to $N$ fermionic flavors as in \cite{Han:2025eiw},

\begin{equation}
\LL =\sum_{a=1}^N\sum_{\sigma=\uparrow,\downarrow} \bar{\psi}_{a,\,\sigma}\Big(\partial_{\tau}-\frac{\vec{\partial}^2}{2m}-\mu\Big)\psi_{a,\,\sigma}-\frac{g}{N}\,(\sum_{a=1}^N \bar{\psi}_{a,\uparrow}\bar{\psi}_{a,\downarrow})\,(\sum_{b=1}^N \psi_{b,\downarrow}\psi_{b,\uparrow})\ .
\end{equation}
This model assumes a $SO(N)\times U(1)_e$ continuous global symmetry under which ${\psi}_{a,\uparrow}$ and  $\bar {\psi}_{a,\downarrow}$ transform as $\mathbf{N}_{1}$ while ${\psi}_{a,\downarrow}$ 
and ${\bar \psi}_{a,\uparrow}$ transform as  $\mathbf{\overline{N}}_{1}$. Under $U(1)_e$,
${\psi}_{a,\uparrow}$, ${\psi}_{a,\downarrow}$  have the same charge equal to $-1$, whereas
$\bar {\psi}_{a,\uparrow}$, $\bar {\psi}_{a,\downarrow}$ have charges equal to 1.\footnote{There is also a $U(1)^N$ global symmetry rotating $(\psi_{a,\uparrow},\bar{\psi}_{a,\downarrow})\rightarrow (e^{i\theta_a}\, \psi_{a,\uparrow},e^{-i\theta_a}\,\bar{\psi}_{a,\downarrow})$ for each $a$. This symmetry will not play any role in the following.}

Introducing a Hubbard-Stratonovich field $\Delta$,
one writes the equivalent Lagrangian

\begin{equation}
\LL'=\sum_{a=1}^N\sum_{\sigma=\uparrow,\downarrow} \bar{\psi}_{a,\,\sigma}\Big(\partial_{\tau}-\frac{\vec{\partial}^2}{2m}-\mu\Big)\psi_{a,\,\sigma}-\Delta\,(\sum_{a=1}^N \bar{\psi}_{a,\uparrow}\bar{\psi}_{a,\downarrow})-\bar{\Delta}\,(\sum_{b=1}^N \psi_{b,\downarrow}\psi_{b,\uparrow})+\frac{N}{g}\Delta\bar{\Delta}\,.
\end{equation}
It is worth noting that, more generically, we could have considered an interaction $\Delta_{ab} \,\bar{\psi}_{a,\uparrow}\bar{\psi}_{b,\downarrow}+{\rm h.c.}$ exhibiting a $U(N)\times U(N)$ global symmetry; of which our model is a restriction. This generic class also includes  models that are relevant for describing the so-called Type 1.5 superconductors \cite{Babaev:2004hk}. 

The equation of motion for $\Delta$ and $\bar{\Delta}$ sets

\begin{equation}
\label{delr}
\bar{\Delta}=\frac{g}{N}\,(\sum_{a=1}^N \bar{\psi}_{a,\uparrow}\bar{\psi}_{a,\downarrow})\,,\qquad \Delta=\frac{g}{N}\,(\sum_{b=1}^N \psi_{b,\downarrow}\psi_{b,\uparrow})\, .
\end{equation}
Substituting back into $\LL'$ we recover the original Lagrangian $\LL$.

As described in the introduction, we are interested in exploring the effect of \textit{a priori} irrelevant couplings. 
We now add  a $\psi^8$ deformation.\footnote{Here we will not consider $\psi^6$ terms because there is no invariant $\psi^6$ interaction  contributing to the $\psi_{a,\downarrow}\psi_{a,\uparrow}$  condensation.}
This term is non-vanishing for $N\geq 2$.
The Lagrangian is
\begin{align}
\label{lagmodel}
 \LL =& \sum_{a=1}^N\sum_{\sigma=\uparrow,\downarrow} \bar{\psi}_{a,\,\sigma}\Big(\partial_{\tau}-\frac{\vec{\partial}^2}{2m}-\mu\Big)\psi_{a,\,\sigma}-\frac{g}{N}\,(\sum_{a=1}^N \bar{\psi}_{a,\uparrow}\bar{\psi}_{a,\downarrow})\,(\sum_{b=1}^N \psi_{b,\downarrow}\psi_{b,\uparrow})\nonumber \\ &  - \frac{\lambda}{2N^3}\,(\sum_{a=1}^N \bar{\psi}_{a,\uparrow}\bar{\psi}_{a,\downarrow})^2(\sum_{b=1}^N \psi_{b,\downarrow}\psi_{b,\uparrow})^2 \,.
\end{align}
where $\lambda \geq 0$ for stability. 
Next, we introduce auxiliary fields $\Delta$ and $\Phi $ as follows

\begin{align}
\label{Lquadratic}
 \LL' = & \sum_{a=1}^N\sum_{\sigma=\uparrow,\downarrow} \bar{\psi}_{a,\,\sigma}\Big(\partial_{\tau}-\frac{\vec{\partial}^2}{2m}-\mu\Big)\psi_{a,\,\sigma}-\Delta\,(\sum_{a=1}^N \bar{\psi}_{a,\uparrow}\bar{\psi}_{a,\downarrow})-\bar{\Delta}\,(\sum_{b=1}^N \psi_{b,\downarrow}\psi_{b,\uparrow})+\frac{N}{g}\Delta\bar{\Delta} \nonumber \\ & -\lambda (\bar \Phi\Phi)\left( \bar\Phi \,(\sum_{a=1}^N \psi_{a,\downarrow}\psi_{a,\uparrow})+\Phi(\sum_{a=1}^N\bar{\psi}_{a,\uparrow}\bar{\psi}_{a,\downarrow})\right)+ \frac{3}{2}\lambda N (\bar\Phi \Phi)^{2}  \,.
\end{align}
The equations of motion for $\bar\Phi$ and  $\Phi$ set
\begin{equation}
\label{Phisol}
 \Phi=\frac{1}{N}\, \left(\sum_{a=1}^N \psi_{a,\downarrow}\psi_{a,\uparrow}\right)
 \ ,\qquad  \bar \Phi=\frac{1}{N}\, \left(
\sum_{a=1}^N\bar{\psi}_{a,\uparrow}\bar{\psi}_{a,\downarrow}\right)\,.
\end{equation}
%
It should be stressed that \cref{lagmodel}  is only one out of many possible models, as one could consider other $\psi^n$ interactions.
Also, for generic $N$, typically there exist various possible contractions of the flavor indices contributing to a given $\psi^n$ interaction. Due to the Grassmann nature of the fermionic field, for sufficiently large $n$, the interaction terms vanish. For instance, for $N=1$ all $\psi^n$ interactions with $n>4$ vanish. For $N=2$ all  $\psi^n$ terms with $n>8$ vanish and  the $\psi^8$ interaction is unique, given by
$$
-\frac{\lambda}{8}(\bar{\psi}_{1,\uparrow}\bar{\psi}_{1,\downarrow})\,( \psi_{1,\downarrow}\psi_{1,\uparrow})(\bar{\psi}_{2,\uparrow}\bar{\psi}_{2,\downarrow})\,( \psi_{2,\downarrow}\psi_{2,\uparrow})\ .
$$
For $N=2$, the Lagrangian \eqref{lagmodel}  represents the most general Lagrangian consistent with the symmetries.

\subsection{Integrating out the fermions}

Writing the model as in \cref{Lquadratic}, the fermions appear quadratically and hence can be integrated out. Let us introduce

\begin{equation}
\Psi_a=\left(\begin{array}{c} \psi_{a,\uparrow}\\ \bar{\psi}_{a,\downarrow}
\end{array}\right)\,,\qquad \Psi^{\dagger}_a=( \bar{\psi}_{a,\uparrow},\,\psi_{a,\downarrow})\,.
\end{equation}
and

\begin{equation}
\mathcal{A}=\left(\begin{array}{cc} -\partial_{\tau} +\frac{\vec{\partial}^2}{2m}+\mu & \Delta+\lambda(\bar\Phi\Phi)\Phi \\ \bar{\Delta}+\lambda(\bar\Phi\Phi)\bar\Phi & -\partial_{\tau} -\frac{\vec{\partial}^2}{2m}-\mu\end{array}\right)\,.
\end{equation}

Then \cref{Lquadratic} becomes

\begin{equation}
 \LL= \sum_{a=1}^N\Psi_a^{\dagger}\,\mathcal{A}\,\Psi_a +\frac{N}{g}\Delta\bar{\Delta} +\frac{3}{2}\lambda N (\bar \Phi \Phi)^{2}  \,.
\end{equation}
Integrating out the $N$ fermions, the one-loop effective Lagrangian becomes


\begin{equation}
\label{Loneloop}
 \LL_{\rm 1-loop}=N\, \left(-{\rm Tr}\log \mathcal{A}+\frac{1}{g}\Delta\bar{\Delta} +\frac{3}{2}\lambda (\bar \Phi \Phi)^{2}\right)\ .
\end{equation}
For the theory at finite temperature, ${\rm Tr}\log \mathcal{A}$ is given by
\begin{equation}
{\rm Tr}\log \mathcal{A}=T\,\sum_n\int \frac{d^d\vec{k}}{(2\pi)^d}\,\log\big[ \big((2n+1)\pi T\big)^2+\xi^2+ |\Delta+\lambda(\bar\Phi\Phi)\Phi|^2\big]\,,
\end{equation}
where
\begin{equation}
\xi=\frac{\vec{k}^2}{2m}-\mu\,.
\end{equation}
Note that ${\rm Tr}\log\mathcal{A}=\mathcal{D}(X)$, with $X=|\Delta+\lambda(\bar\Phi\Phi)\Phi|^2$. 
The equations of motion for $\bar \Delta$, $\Delta $ are
\begin{equation}
\label{deltaeqq}
 -[\Delta+\lambda(\bar\Phi\Phi)\Phi]\, \mathcal{D}'(X)+\frac{\Delta}{g}=0\, \qquad -[\bar \Delta+\lambda(\bar\Phi\Phi)\bar \Phi]\, \mathcal{D}'(X)+\frac{\bar \Delta}{g}=0\,.
\end{equation}
where we have introduced the notation $\mathcal{D}'\equiv \partial \mathcal{D}/\partial X$. 
Let us now consider the equations for  $\bar\Phi,\ \Phi$.
In view of \eqref{delr}, \eqref{Phisol}, one can anticipate that the solution will
be $\bar\Phi=\bar\Delta/g$\ , $\Phi=\Delta/g$
As usual, one can set  $\bar \Delta=\Delta$ (the constant phase
is irrelevant for the dynamics as it can be removed by a $U(1)$ rotation of the fermions).
This sets  $\bar\Phi =\Phi$. The remaining equation for $\Phi $ is

\begin{equation}
\label{phieqq}
 -\Phi^2 [\Delta+\lambda\, \Phi^3]\, \mathcal{D}'(X)+ \Phi^3=0\,.
\end{equation}
There is a trivial solution $\Delta=\Phi=0$ representing the uncondensed phase. To look for the condensed phase, we assume $\Delta\neq 0$ and $\Phi\neq 0$. Then,
combining \eqref{deltaeqq} and \eqref{phieqq}, we find, as expected 
\begin{equation}
\label{dff}
\Phi=\frac{1}{g}\, \Delta\,.
\end{equation}
We are finally then left with a single equation, which reads

\begin{equation}
\label{gapeq1}
\left(-(1+c\,\Delta^2)\,\mathcal{D}'+\frac{1}{g}\right)\Delta =0\,.
\end{equation}
where 
\begin{equation}
    c\equiv\frac{\lambda}{g^3}\ ,
\end{equation}
and $\mathcal{D}'$ is evaluated at $\Phi$
given by \eqref{dff}. 

 The gap  in the condensed phase is determined by
the   equation
\begin{equation}
(1+c\,\Delta^2)\, \mathcal{D}'=\frac{1}{g}\ .
\end{equation}
In order to further proceed, we now need to compute $\mathcal{D}'(X)$.
Explicitly, it reads
\begin{equation}
\mathcal{D}'=T\,\sum_n\, \int \frac{d^d\vec{k}}{(2\pi)^d}\, \frac{1}{\big((2n+1)\pi T\big)^2+\xi^2+ \Delta ^2\,\eta^2}\,,
\end{equation}
where 
\begin{equation}
    \eta\equiv 1+c\,\Delta^2\ .
\end{equation}
Performing the sum, one finds

\begin{equation}
\mathcal{D}'=\frac{1}{2}\, \int \frac{d^d\vec{k}}{(2\pi)^d}\,\frac{\tanh\frac{\sqrt{\xi^2+ \Delta^2\,\eta^2}}{2T}}{\sqrt{\xi^2+ \Delta^2\,\eta^2}}\,.
\end{equation}
Integrating, we find

\begin{equation}
\mathcal{D}=\int \frac{d^d\vec{k}}{(2\pi)^d} \Big[ \sqrt{\xi^2+ \Delta^2\,\eta^2}+2T\,\log\frac12\Big( 1+e^{-\frac{\sqrt{\xi^2+ \Delta^2\,\eta^2}}{T}}\Big)\Big]\,.
\end{equation}
This formula will be used in the calculation of the free energy.

In standard BCS theory, one assumes that the relevant physics takes place near the Fermi surface. Under this assumption, one can approximate 

\begin{equation}
\label{momentumaproximation}
\int \frac{d^d\vec{k}}{(2\pi)^d}\sim \nu\,\int_{-\omega_*}^{\omega_{\star}}d\xi\,,
\end{equation}
where $\omega_{\star}$ is the Debye energy (playing the role of a physical cutoff) and $\nu $ is the density of states at the Fermi level, defined by a $\xi(k) =\frac{k^2}{2m}-\mu=0$. The final form of the gap equation is therefore:

\begin{equation}
\label{ecugap}
 \frac{1}{g\,\nu\,(1+c\,\Delta^2)}= \int_0^{\omega_{\star}}d\xi\,\frac{\tanh\frac{\sqrt{\xi^2+  \Delta^2\,\eta^2}}{2T}}{\sqrt{\xi^2+ \Delta^2\,\eta^2}}\,.
\end{equation}
The new term with coefficient  $c$ represents a deviation from standard BCS theory and cannot be absorbed into a redefinition of $\Delta$.

\subsection{Free energy}

The free energy is obtained by computing the action evaluated on the solution. This yields (we set $\Delta=\bar{\Delta}$)

\begin{equation}
\frac{F}{N}=\frac{1}{g}\Delta^2+\frac{3}{2}\lambda  \Phi^{4}
-\int \frac{d^d\vec{k}}{(2\pi)^d}\Big[ \sqrt{\xi^2+ \Delta^2\,\eta^2}+2T\,\log\frac12\Big( 1+e^{-\frac{\sqrt{\xi^2+ \Delta^2\,\eta^2}}{T}}\Big)\Big]\,.
\end{equation}
Choosing the uncondensed state --with free energy $F_0$-- as the reference, the free-energy difference $ f\equiv \frac{1}{N}(F-F_0)$ is

\begin{equation}
f=\frac{1}{g}\Delta^2 +\frac{3\, c}{2\,g}\,\Delta^4-\int \frac{d^d\vec{k}}{(2\pi)^d}\Big[ \big(\sqrt{\xi^2+ \Delta^2\,\eta^2}-\xi\big)+2T\,\log\Big( \frac{1+e^{-\frac{\sqrt{\xi^2+ \Delta^2\,\eta^2}}{T}}}{1+e^{-\frac{\xi}{T}}}\Big)\Big]\,,
\end{equation}
Assuming the same approximation as in \cref{momentumaproximation} for the momentum integral we find

\begin{equation}
\label{freeequ}
 f=\frac{1}{g}\Delta^2 + \frac{3\,c}{2\,g}\,\Delta^4-2\nu \int_0^{\omega_{\star}} d\xi \Big[ \big(\sqrt{\xi^2+ \Delta^2\,\eta^2}-\xi\big)+2T\,\log\Big( \frac{1+e^{-\frac{\sqrt{\xi^2+ \Delta^2\,\eta^2}}{T}}}{1+e^{-\frac{\xi}{T}}}\Big)\Big]\,.
\end{equation}

\subsection{Fluctuations}

To analyze the system we have so far used the saddle-point method applied to the one-loop action \eqref{Loneloop}. Owing to the overall $N$ factor, this approximation becomes exact in the infinite $N$ limit. In order to  examine the stability, we can study the fluctuations around the classical configuration. Writing the fields as $\Delta=\Delta_0+\delta \Delta$, $\Phi=\Phi_0+g^{-1}\delta \Phi$, where $\Delta_0$, $\Phi_0$ are the saddle-point values of the fields, one can systematically expand \eqref{Loneloop} in perturbation theory around the classical solution. To quadratic order, we can read off the mass matrix for the fluctuations. It is straightforward to check that such matrix has one zero eigenvalue, corresponding to the Goldstone mode arising from the spontaneous symmetry breaking of the $U(1)_e$. The rest of the eigenvalues are positive, corresponding to positive squared masses for the fluctuations. The explicit expressions are rather lengthy. For $c\ll 1$, to linear order in $c$, they simplify and read

\begin{equation}
M^2=\{0,\, \frac{c\,\Delta_0^2}{g},\, -\mathcal{D}''\,\Delta_0^2+\frac{c\,\Delta_0^2}{2\,g}(1-4\,\mathcal{D}''\,g\,\Delta_0^2),\,\frac{3\,c\,\Delta_0^2}{2\,g}\}\,,
\end{equation}
where 
\begin{equation}
    \mathcal{D}''(X)=-\frac{\nu}{4}\int_0^{\omega_\star}\ d\xi\  
    \frac{
    \sinh\frac{\sqrt{\xi^2+X}}{T} -\frac{\sqrt{\xi^2+X}}{T}}{ (\xi^2+X)^{\frac32 } \cosh^2\frac{\sqrt{\xi^2+X}}{2T}}\ .
\end{equation}
 $\mathcal{D}''$ is given by minus the integral of a positive definite quantity.
Therefore $\mathcal{D}''$ is negative,
$\mathcal{D}''<0$, for any value of $X$.
Thus, apart from the protected massless Goldstone mode, all fluctuations around the vacuum possess positive squared mass, ensuring the perturbative stability of the model.

\section{Analysis of the condensed phase}\label{analysiscondensed}

In this section, an analytical study of the possible condensed phases is carried out by examining the gap equation \cref{ecugap}.

\subsection{The $T=0$ limit}

We first consider the small temperature region. As $T\rightarrow 0$, the $\tanh$ in \cref{ecugap} goes to one, and the integral can be computed explicitly. One finds the following  equation for $\Delta_0\equiv\Delta(T=0)$:
\begin{equation}
    \eta_0\, \Delta_0 =\frac{\omega_{\star}}{\sinh(1/(g\, \nu\, \eta_0)) }\ ,\qquad  \eta_0\equiv 1+c\Delta_0^2\ .
\end{equation}
This is a transcendental equation for $\Delta$
that can be solved numerically. 
The BCS
 case, obtained by setting
$c=0$, gives an estimate, which also applies to case where $c$ is small,
\begin{equation}
\label{demax}
    \Delta_0\approx\frac{\omega_{\star}}{\sinh(1/g\nu)}\ , \qquad {\rm for}\ \ c\, \Delta_0^2\ll 1\ .
\end{equation}

\subsection{Expansion near the critical temperature}

Assuming that $\Delta(T)$ vanishes at a critical temperature
$T_c$, we have
\begin{equation}
\frac{1}{g\nu}= \int_0^{\frac{\omega_{\star}}{T_c}} \frac{\tanh\frac{x}{2}}{x}\approx \log \frac{k \omega_{\star}}{T_c}\,,\quad k\approx 1.134\qquad \leadsto \qquad T_c\approx  k\, \omega_{\star} e^{-\frac{1}{g\nu}}\ .
\end{equation}
Note that the critical temperature $T_c$ at which the gap vanishes, $\Delta=0$, is  the same as in BCS theory since in this case the correction with  $c$ coefficient vanishes. However, because $\Delta $ can in principle be multivalued,
the  temperature $T_c$ may not be the critical temperature of the onset of the phase transition.
In  section \ref{numerical} we will examine instances in which the system undergoes a first-order phase transition at a higher critical temperature $T_c^{\star}$.

The region close to $T_c$ is amenable to analytical treatment.  
Let us expand the gap equation around  $T_c$. Defining $\delta T=T_c-T$ and neglecting $O(\Delta^3)$ terms, we get

\begin{equation}
\alpha_1\,\delta T-\left(\beta_0-\frac{c}{g\nu}\right)\,\Delta^2=0\,.
\end{equation}
where

\begin{equation}
 \alpha_1=\frac{1}{2T_c}\, \int_0^{\frac{\omega_{\star}}{T_c}} dx\, \frac{1}{\cosh^2\frac{x}{2}}\,,\qquad 
\beta_0= \frac{1}{4T_c^2}\, \int_0^{\frac{\omega_{\star}}{T_c}} dx\  \frac{1}{x^3\,\cosh^2\frac{x}{2}}\,\left(\sinh(x)-x\right)\,.
\end{equation}

It follows that
\begin{equation}
\Delta^2=\frac{\alpha_1}{\beta_0-\frac{c}{g\nu}}\,\delta T\,.
\end{equation}
We recall that $c\geq 0$ for stability of the potential. 
Thus we have a second order transition provided $\beta_0-\frac{c}{g\nu}>0$, and first order otherwise. The critical $c$ happens at $c_0=g\nu\,\beta_0$. Explicitly

\begin{equation}
c_0=\frac{1}{4T_c^2}\, \frac{ \int_0^{\frac{\omega_{\star}}{T_c}}  \frac{1}{x^3\,\cosh^2\frac{x}{2}}\,(\sinh x-x)}{ \int_0^{\frac{\omega_{\star}}{T_c}} \frac{\tanh\frac{x}{2}}{x}}\approx n_0\, \frac{g\nu}{4T_c^2},\ \ n_0\approx 0.426\,,
\end{equation}
where the integral in the numerator has been estimated assuming that 
$\omega_{\star}\gg T_c$. In order to study the behavior right at the critical coupling $c=c_0$, 
we need to keep the next order in the gap $\Delta(T)$, leading to
\begin{equation}
 \Delta^4 =\gamma\, (T_c-T)+ O(\delta T^2)\ ,\qquad \gamma>0\ .
\end{equation}
Thus, at this critical coupling $c_0$, the critical exponent is $1/4$.

To provide further support for this picture, we can calculate the free energy near $T_c$. We find

\begin{align}
\label{fttt}
    f\sim \frac{1}{2}\,\frac{\alpha_1^2\,\nu}{(\beta_0-\frac{c}{g\nu})}\,\delta T^2+\cdots \,.
\end{align}
Thus we see that $f$ precisely changes sign at $c=c_0$ and goes from a negative to a positive concavity at $T_c$, consistent with the change from a second to a first order phase transition (\textit{cf.} red, black and blue \textit{vs.} green curves in \cref{fig:tau} below).

Just as in conventional BCS (type I) superconductors, the Meissner effect dictates the temperature dependence of the critical magnetic field. One has
\begin{equation}
    \frac12 H_c(T)^2 = N f(T)\ .
\end{equation}
For $b=0$ and $c<c_0$,
near the critical temperature, using \eqref{fttt}, one obtains the usual behavior
\begin{equation}
H_c(T)\approx H_c(0)\left(1-\left(\frac{T}{T_c}\right)^2\right)\ . 
\end{equation}
Away from the critical temperature, in
regions where $\Delta $ is not small, there will be deviations 
due to the influence of the $\psi^8$ interaction, but the phase diagram  will  be qualitatively the same.

\section{Numerical analysis of the condensed phase}\label{numerical}

In this section we numerically solve  the gap equation and compute the free energy of the system for different values of the parameters.

\begin{figure}[h!]
 \centering
\includegraphics[width=0.5\textwidth]{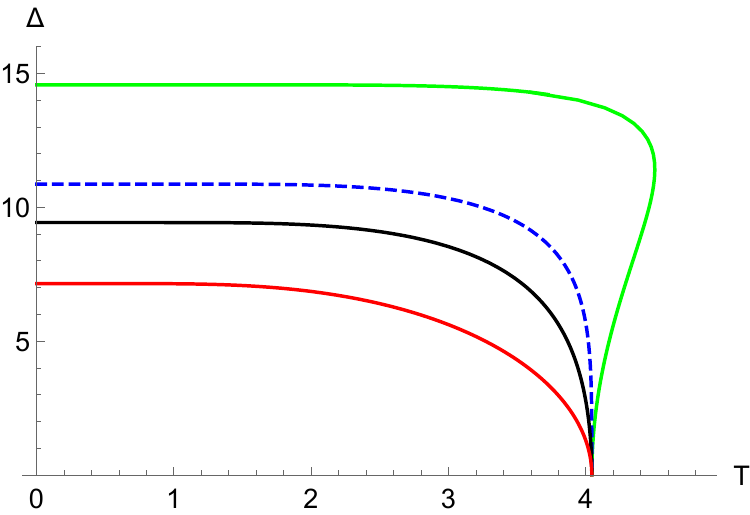}
  \caption{$\Delta$ \textit{vs.} $T$ for $c=0$ (standard BCS, in red), then $c=0.8\, c_0$ (black), $c=c_0$ (dashed, blue), $c=1.2\, c_0$ (green). Here $g\nu=0.3$, $\omega_{\star}=100$.}
 \label{fig1}
\end{figure}

 According to the analysis of the previous section, the
theory with $c<c_0$ should exhibit a second order phase transition with critical exponent $1/2$ ($1/4$ for the fine-tuned case $c=c_0$) and a first order phase transition in the regime $c>c_0$. This is supported by \cref{fig1}, where we show the numerical solution to the gap equation for $\Delta$ as a function of the temperature for different values of $c$. For $c<c_0$ (red and black  curves) the curve has a similar behavior to the standard BCS case. In turn, for $c>c_0$ (green curve), $\Delta$ develops a lump (in the case at hand, reaching up to $T_0\sim 4.5$) and becomes non-single valued in the range $T\in (T_c,\,T_0)$, indicating the presence of a  first order phase transition in this regime.

\begin{figure}[h!]
 \centering
 \begin{tabular}{cc}
\includegraphics[width=0.4\textwidth]{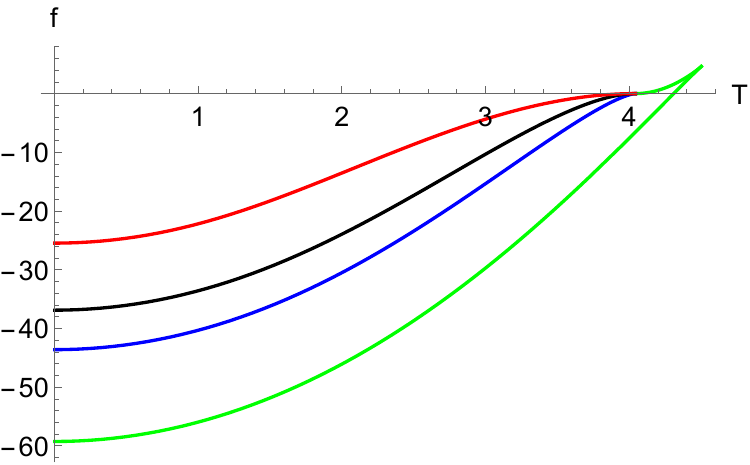}
 &
 \qquad \includegraphics[width=0.4\textwidth]{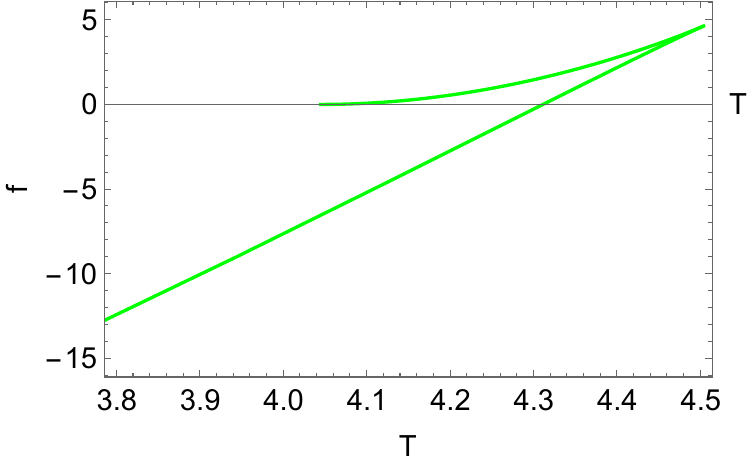}
 \\ (a)&(b)
 \end{tabular}
 \caption{a)  Free energy \textit{vs.} $T$ (same conventions as in figure 1).
 b) Enlarged view of the free energy in the case $c=1.2\, c_0$. The lower branch in Fig. 1 corresponds to the branch with positive free energy. The upper branch in figure 1 has negative free energy until  $T^{\star}_{c}\approx 4.3$, where a first-order phase transition takes place. For $T>T^{\star}_{c}$, $\Delta=0$.
 }
 \label{fig:tau}
 \end{figure}

This scenario can be explicitly tested by computing the free energy in each case as shown in \cref{fig:tau}. For $c>c_0$ the free energy exhibits the characteristic swallowtail of a first-order phase transition showing that at some $T_c^{\star}$ ($\sim 4.3$ in the example shown) there is a first order phase transition.

 The critical exponents in the region $c < c_0$ remain the same as in  mean-field theory independent of the value of $c$.
The temperature dependence of the gap exhibits a qualitatively distinct profile as compared to the $c=0$ case (\textit{cf.} the black curve with the red curve in \cref{fig1}).

\section{Conclusions}\label{conclusions}

In this paper we have studied the effect of  $\psi^4$ and $\psi^8$ interactions interactions at finite temperature and finite chemical potential in the context of BCS-like phase transitions. While these interactions might naively be expected to play a minor role because these operators are ``superficially" irrelevant operators, the present results indicate that they can significantly affect the system’s behavior, in some instances significantly modifying temperature dependence of the gap  or leading to a first-order transition.
The mechanism by which an irrelevant operator turns into a marginally relevant operator is well understood in BCS theory \cite{Polchinski:1992ed}.
In the standard BCS theory, the effective quartic interaction is generated through electron-phonon interactions.  Below a critical temperature, the quartic interaction forces fermions to condense into Cooper pairs driving superconductivity. 
Under naive dimensional analysis, the quartic coupling possesses a mass dimension of $2-d$, rendering it irrelevant in spacetime dimensions $d>2$.
Thus, from the standpoint of effective field theory, it may be unexpected that a quartic fermionic interaction can lead to condensation. 
Nevertheless, as shown in \cite{Polchinski:1992ed}, the presence of a Fermi surface fundamentally modifies this scaling, rendering the quartic interaction marginally relevant within particular kinematic regimes. This observation naturally prompts a systematic investigation into the behavior of higher-order couplings and their potential role in mediating analogous effects.
Although a comprehensive investigation lies beyond the scope of this work, in Section 1 we presented a schematic example illustrating how the RG flow intermixes the evolution of the different couplings,
thereby propagating the low-energy behavior of the quartic fermion coupling in a way that can substantially affect the low-energy behavior of 
the $\psi^8$ fermionic interaction.


\smallskip

When the $\psi^8$ interaction is turned off, the model describes  BCS theory with $N$ fermion species and the usual $\psi^4$ interaction. In practical applications, models with several species of fermions arise in multiband  superconductors \cite{kruchinin}, where fermions of different bands are treated as different species.
Higher-order fermionic interactions are expected to arise quite generically in the effective Lagrangian whenever the system involves multiple components, i.e., for $N \geq 2$. This observation is particularly relevant in the context of multicomponent superconductors, and in particular type-1.5 systems, which have attracted considerable attention in recent years (see \cite{Babaev:2004hk}  for a comprehensive review). Type-1.5 superconductors typically possess at least two distinct superconducting condensates, and their phenomenology can be captured within a generalized Ginzburg--Landau framework that incorporates $N=2$ fermionic degrees of freedom coupled to a dynamical vector potential. In the superconducting phase, the vector potential acquires a finite mass through the Anderson-Higgs mechanism, and its propagator mediates interactions that generate nontrivial Feynman diagrams contributing, in particular, to  $\psi^6$ and $\psi^8$ operators. As emphasized in the introduction, such higher-order terms are not only induced by gauge-field exchange but also emerge from fermion loop corrections associated with the quartic $\psi^4$ interaction.  While in many conventional superconductors these contributions are suppressed and can be safely neglected, their role may become nontrivial in type-1.5 systems characterized by strong electron--phonon coupling. In such regimes, higher-order fermionic interactions can significantly modify the low-energy effective theory, potentially leading to observable deviations from standard BCS behavior, including altered vortex dynamics, modified coherence lengths, and unconventional collective excitations. 

\smallskip

The temperature dependence of the gap in type‑1.5 superconducting systems is typically flatter for most of the temperature range and drops sharply near $T_c$ (see e.g. \cite{khasanov}).
This is in accordance with the behavior exhibited by the models with  $c<c_0$ (black curve in Fig. \ref{fig1}).
Given this remarkable agreement, it would be highly interesting to investigate whether strong electron-phonon coupling underlies the effective $\psi^ 8$ interaction. A more fundamental derivation of the 
$\psi^ 8$ interaction could also provide valuable insight into estimating the coupling constant $c$ in these superconducting systems.

\smallskip

For $c>c_0$, the gap becomes double valued above $T_c$.
This feature appears in a number of models, in particular, in the Eliashberg-Nambu theory for strong-coupling superconductors \cite{Zheng,Marsiglio,Marsiglio2}.
In this parameter range, the present models exhibit a first-order superconducting transition, where the gap edge drops from a finite value to zero.
While most common superconductors have a continuous superconducting transition,
there are cases where a discontinuous jump in the order parameter is expected.
This includes multicomponent superconductors, strong-coupling Eliashberg superconductors and heavy fermions (first-order phase transitions also appear in the presence of magnetic fields).

\medskip

\section*{Acknowledgements}

 JGR acknowledges financial support from the Spanish  MCIN/AEI/10.13039/501100011033 grant PID2022-126224NB-C21. DRG is grateful to the University of Barcelona for hospitality. DRG is supported in part by the Spanish national grant MCIU-22-
PID2021-123021NB-I00

\end{document}